\documentclass[aps,preprintnumbers,nofootinbib,twocolumn, superscriptaddress]{revtex4-1}

\usepackage{graphicx}
\usepackage[shortlabels]{enumitem}
\usepackage{mathtools}
\usepackage{nicefrac}
\usepackage{mathrsfs}
\usepackage{subcaption}
\usepackage{caption}
\usepackage{bbm}
\usepackage{upgreek}
\usepackage{makecell}
\usepackage{placeins}
\usepackage{afterpage}
\usepackage{titlesec}
\usepackage{amsfonts,amssymb,amsmath}
\usepackage{ifthen}
\usepackage{amsthm, thm-restate}
\usepackage{dsfont}
\usepackage{float}
\usepackage{MnSymbol}
\usepackage{cancel}
\PassOptionsToPackage{hyphens}{url}
\usepackage[hyperfootnotes=false]{hyperref}
\usepackage{xcolor}
\definecolor{mylinkcolor}{rgb}{0,0,0.7} % set link color here as red,green,blue.
\hypersetup{unicode=true,%
  bookmarksnumbered=false,bookmarksopen=false,bookmarksopenlevel=1, %
  breaklinks=true,pdfborder={0 0 0},colorlinks=true}%
\hypersetup{%
  anchorcolor=mylinkcolor,citecolor=mylinkcolor, %
  filecolor=mylinkcolor,linkcolor=mylinkcolor, %
  menucolor=mylinkcolor,runcolor=mylinkcolor, %
  urlcolor=mylinkcolor}

\usepackage[capitalise]{cleveref}

\usepackage{tikz}
\usetikzlibrary{arrows}
\usetikzlibrary{arrows.meta}
\usetikzlibrary{shapes}
\usetikzlibrary{hobby}
\usetikzlibrary{decorations.markings}
\usetikzlibrary{calc,intersections,through,backgrounds}
\usetikzlibrary{decorations.pathreplacing,angles,quotes}
\usetikzlibrary{arrows.meta}
\usetikzlibrary{snakes}

\theoremstyle{plain}
\newtheorem{theorem}{Theorem}

\newtheorem{corollary}{Corollary}

\theoremstyle{definition}
\newtheorem{definition}{Definition}[section]

    \newcommand{\ket}[1]{\vert  #1 \rangle}
    \newcommand{\bra}[1]{\langle #1 |}

	\newcommand{\tr}{\operatorname{Tr}  }

\usepackage[nodayofweek]{datetime}
\newcommand{\comment}[1]{}

\newcommand{\cE}{\mathcal{E}}
\newcommand{\cF}{\mathcal{F}}
\newcommand{\cG}{\mathcal{G}}
\newcommand{\cH}{\mathscr{H}}
\newcommand{\cI}{\mathcal{I}}
\newcommand{\cL}{\mathcal{L}}
\newcommand{\cM}{\mathcal{M}}
\newcommand{\cN}{\mathcal{N}}
\newcommand{\cP}{\mathcal{P}}
\newcommand{\cQ}{\mathcal{Q}}
\newcommand{\cR}{\mathcal{R}}
\newcommand{\cS}{\mathcal{S}}
\newcommand{\cT}{\mathcal{T}}
\newcommand{\cW}{\mathcal{W}}

\makeatletter
\pgfkeys{/pgf/.cd,
  cube offset x/.initial=2mm,
  cube offset y/.initial=2mm
}
\pgfdeclareshape{cube}
{
  \inheritsavedanchors[from=rectangle] % this is nearly a rectangle
  \inheritanchorborder[from=rectangle]
  \inheritanchor[from=rectangle]{north}
  \inheritanchor[from=rectangle]{north west}
  \inheritanchor[from=rectangle]{north east}
  \inheritanchor[from=rectangle]{center}
  \inheritanchor[from=rectangle]{west}
  \inheritanchor[from=rectangle]{east}
  \inheritanchor[from=rectangle]{mid}
  \inheritanchor[from=rectangle]{mid west}
  \inheritanchor[from=rectangle]{mid east}
  \inheritanchor[from=rectangle]{base}
  \inheritanchor[from=rectangle]{base west}
  \inheritanchor[from=rectangle]{base east}
  \inheritanchor[from=rectangle]{south}
  \inheritanchor[from=rectangle]{south west}
  \inheritanchor[from=rectangle]{south east}
  \backgroundpath{
    \southwest \pgf@xa=\pgf@x \pgf@ya=\pgf@y
    \northeast \pgf@xb=\pgf@x \pgf@yb=\pgf@y
    \pgfmathsetlength\pgfutil@tempdima{\pgfkeysvalueof{/pgf/cube offset x}}
    \pgfmathsetlength\pgfutil@tempdimb{\pgfkeysvalueof{/pgf/cube offset y}}
    \def\ppd@offset{\pgfpoint{\pgfutil@tempdima}{\pgfutil@tempdimb}}
    \pgfpathmoveto{\pgfqpoint{\pgf@xa}{\pgf@ya}}
    \pgfpathlineto{\pgfqpoint{\pgf@xb}{\pgf@ya}}
    \pgfpathlineto{\pgfqpoint{\pgf@xb}{\pgf@yb}}
    \pgfpathlineto{\pgfqpoint{\pgf@xa}{\pgf@yb}}
    \pgfpathclose
    \pgfpathmoveto{\pgfqpoint{\pgf@xb}{\pgf@ya}}
    \pgfpathlineto{\pgfpointadd{\pgfpoint{\pgf@xb}{\pgf@ya}}{\ppd@offset}}
    \pgfpathlineto{\pgfpointadd{\pgfpoint{\pgf@xb}{\pgf@yb}}{\ppd@offset}}
    \pgfpathlineto{\pgfpointadd{\pgfpoint{\pgf@xa}{\pgf@yb}}{\ppd@offset}}
    \pgfpathlineto{\pgfqpoint{\pgf@xa}{\pgf@yb}}
    \pgfpathmoveto{\pgfqpoint{\pgf@xb}{\pgf@yb}}
    \pgfpathlineto{\pgfpointadd{\pgfpoint{\pgf@xb}{\pgf@yb}}{\ppd@offset}}
  }
}
\makeatother

\newcommand\longrsquigarrow{
\begin{tikzpicture}
\draw [decorate, decoration={zigzag, segment length=+6pt, amplitude=+.95pt,post length=+2pt}, arrows={-Classical TikZ Rightarrow}]  (0,0.1) -- (0.6,0.1); \draw[draw=none] (0,0)--(0.6,0);
\end{tikzpicture}
}

\DeclareRobustCommand\ground{\begin{tikzpicture}[scale=0.8]
		\draw [thick](-0.4,0)--(0.4,0);\draw [thick](-0.3,0.1)--(0.3,0.1);\draw [thick](-0.2,0.2)--(0.2,0.2);\draw [thick](-0.1,0.3)--(0.1,0.3);
	\end{tikzpicture}}

\newpage

\begin{document}

%\title{No-go theorems for reconciling information-theoretic and spatiotemporal causality}
\title{Fundamental limits for realising quantum processes in spacetime}

\author{V. Vilasini}
\email{vilasini@inria.fr}
\affiliation{Université Grenoble Alpes, Inria, 38000 Grenoble, France}
\affiliation{Institute for Theoretical Physics, ETH Zurich, 8093 Z\"{u}rich, Switzerland}
\author{Renato Renner}
\email{renner@ethz.ch}
\affiliation{Institute for Theoretical Physics, ETH Zurich, 8093 Z\"{u}rich, Switzerland}

\date{\today}

\begin{abstract}

Understanding the interface between quantum and relativistic theories is crucial for fundamental and practical advances, especially given that key physical concepts such as causality take different forms in these theories. Bell's no-go theorem reveals limits on classical processes, arising from relativistic causality principles. Considering whether similar fundamental limits exist on quantum processes, we derive no-go theorems for quantum experiments realisable in classical background spacetimes. We account for general processes allowed by quantum theory, including those with indefinite causal order (ICO), which have also been the subject of recent experiments. Our first theorem implies that realisations of ICO processes that do not violate relativistic causality must involve the non-localization of systems in spacetime. The second theorem shows that for any such realisation of an ICO process, there exists a more fine-grained description in terms of a definite and acyclic causal order process. This enables a general reconciliation of quantum and relativistic notions of causality and, in particular, applies to experimental realisations of the quantum switch, a prominent ICO process. By showing what is impossible to achieve in classical spacetimes, these no-go results also offer insights into how causality and information processing may differ in future quantum experiments in relativistic regimes beyond classical spacetimes.

\end{abstract}

\maketitle

\noindent{\it Introduction.}| Understanding the interface between quantum and relativistic theories is imperative in the fundamental context of quantum gravity and for developing relativistic quantum information technologies. Concepts such as causality and locality, which are central to how we make sense of the world, take on different forms in these theories (see also \cite{VilasiniColbeckPRA, VilasiniColbeckPRL, us_long}).%, and their relations are not completely understood. %In relativistic physics, these concepts are linked to the geometry of space and time, while in information-theoretic domains like classical statistics and quantum theory, they relate to the flow of information. The relation between these causality notions is not completely understood, although they play together compatibly in physical experiments. 

Bell's seminal no-go theorem \cite{Bell1964} reveals fundamental limits on classical processes. From minimal axioms such as relativistic causality principles in spacetime (the impossibility of superluminal causation) and free choice, it shows that such classical processes cannot explain correlations observed in quantum experiments \cite{Wiseman2015}. This indicates that causality must work differently in the quantum world in order to preserve free choice and compatibility with relativistic principles \cite{Note1}, fuelling a growing research program on \emph{quantum information-theoretic approaches to causality} (e.g., \cite{Hardy2005, Henson2014, Costa2016, Leifer2013, Barrett2020, Barrett2020A, Oreshkov2012, Chiribella2013}). %Note that no superluminal causation (NSC) and no superluminal signalling (NSS) are distinct relativistic principles \cite{Wiseman2015,VilasiniColbeckPRA,VilasiniColbeckPRL, VilasiniColbeck2024}, the former being stronger than the latter. For instance, deterministic classical processes exist (e.g., in non-local hidden variable theories) that can explain quantum correlations while satisfying NSS but violating NSC. Bell's theorem rules out such classical explanations under the stronger NSC principle (and free choice).

Going further, a natural question is whether similar fundamental limits on quantum processes arise from relativistic principles in spacetime. In this work, we derive no-go theorems to address this question. We focus on classical background spacetimes, which is the regime that is currently experimentally accessible. To be fully general, we consider all processes allowed by quantum theory, which includes indefinite causal order (ICO) quantum processes \cite{Hardy2005, Chiribella2013, Oreshkov2012}, modelling quantum protocols which lack a fixed acyclic order between the quantum operations. ICO processes have been widely studied for the potential advantages they may offer for quantum information processing (e.g., \cite{Guerin2016, Chiribella_2021,Chiribella_2012,Zhao_2020,Araujo2014,Guha_2020,mothe2023}). However, important questions about their physical realisability remain open. %Our general approach will enable us to also shed light on these questions.

Specifically, numerous table-top experiments \cite{Procopio2015, Rubino2017, Goswami2018, Wei_2019, Ho2019, Guo_2020, Goswami_2020, Taddei_2021, Rubino_2021, Felce2021}  have claimed to implement a particular ICO process, the quantum switch \cite{Chiribella2013}, in Minkowski spacetime (a definite relativistic causal structure). There have been longstanding discussions within the community about interpreting these experiments \cite{Portmann2017,Vilasini_thesis,Paunkovic2019, Oreshkov2019, Ormrod2022, Kabel2024}. Our results answer a fundamental question at the core of this discussion: how can an \emph{indefinite quantum causal structure} consistently coexist with a \emph{definite relativistic causal structure}? 

Specifically, our first no-go theorem implies that, to physically realise ICO quantum processes in spacetime, it is necessary for the quantum in/output systems of the operations not to be localised in spacetime. This feature goes beyond Bell-type experiments, where one regards these systems as being well-localised. Our second no-go theorem shows that, as a consequence of relativistic causality, even such realisations will ultimately admit a fine-grained description in terms of a definite and acyclic causal order process compatible with the spacetime structure. %While resolving question \textbf{Q}, these yield a clear physical interpretation to the afore-mentioned experiments. 

To formalise these results rigorously, we propose a general method that connects quantum and relativistic approaches to causality, which we present below. In particular, we introduce the novel concept of fine-graining of quantum networks, which is also applicable to cyclic quantum causal structures and is crucial for reconciling the two causality notions. A comprehensive framework backing this method, with additional results and applications, is presented in our accompanying paper \cite{us_long}. %More generally, our work lays a foundation for investigating how causality and information-processing possibilities may differ in quantum gravitational regimes, where spacetime may no longer be classical and fixed. 

\noindent{\it Cyclic quantum networks}| The standard quantum circuit paradigm describes information-theoretic structures with a definite and acyclic ordering between operations. Here, we start with a more general class of information-theoretic structures, \emph{cyclic quantum networks}, which will allow us to recover indefinite causal order processes (and, hence, their subset, standard circuits) as special instances.

We define a \emph{quantum network} $\mathfrak{N}:=(\mathfrak{N}^{\mathrm{maps}},\mathfrak{N}^{\mathrm{comp}})$ by a set $\mathfrak{N}^{\mathrm{maps}}$ of completely positive maps (CPMs) together with a set $\mathfrak{N}^{\mathrm{comp}}$ of compositions, where each composition, represented as $O \hookrightarrow I$, indicates that the output system $O$ of a CPM is connected to the input system $I$ of another (or the same) CPM in the network. Each in/output system can be involved in at most one composition.
We denote by~$\mathfrak{N}^{\mathrm{sys}}$ the set  of all input systems $I$ and output systems $O$ of the CPMs in~$\mathfrak{N}^{\mathrm{maps}}$, with $O$ and $I$ identified whenever $O \hookrightarrow I \in \mathfrak{N}^{\mathrm{comp}}$. A system is called  \emph{free} if it is not composed with any other systems through $\hookrightarrow$. Furthermore, a \emph{sub-network} $\mathfrak{N}_{\mathrm{sub}}$ of~$\mathfrak{N}$  is a network consisting of a subset of the CPMs and a subset of the set of compositions of $\mathfrak{N}$. By plugging together all the CPMs in $\mathfrak{N}^{\mathrm{maps}}_{\mathrm{sub}}$ through all the compositions in $\mathfrak{N}^{\mathrm{comp}}_{\mathrm{sub}}$, one obtains a map $\mathcal{N}_{\mathrm{sub}}$ from the free inputs to the free outputs of $\mathfrak{N}_{\mathrm{sub}}$. This is called the \emph{induced map} of the (sub)network, a CPM that is well-defined independently of the order in which the compositions are performed.

\noindent{\it Signalling and compatibility}| Given a quantum network, we can operationally detect the influence of one set $A$ of systems on another disjoint set $B$ of systems through \emph{signalling relations}. This involves checking whether two different interventions (quantum operations) on $A$ lead to distinct (distinguishable) states on $B$. The signalling structure $\mathcal{G}^{\mathrm{sig}}_{\mathfrak{N}}$ of a network corresponds to a directed graph with nodes corresponding the powerset $\Sigma(\mathfrak{N}^{\mathrm{sys}})$, and a directed edge $\cS_1\longrightarrow \cS_2$ between sets of systems whenever there is signalling. A useful concept is that of \emph{compatibility} between a signalling structure and another directed graph $\mathcal{G}$ (this may, for instance, capture the causal structure of spacetime). Suppose each system $S\in \mathfrak{N}^{\mathrm{sys}}$ is assigned a node $\mathcal{E}(S)\in \mathrm{Nodes}(\mathcal{G})$ of  $\mathcal{G}$ through some mapping $\mathcal{E}$. Then we say that $\mathcal{G}^{\mathrm{sig}}_{\mathfrak{N}}$ is compatible with $\mathcal{G}$ iff for any two disjoint sets $\mathcal{S}_1$ and $\mathcal{S}_2$ of systems in the network, denoting $\mathcal{E}(\mathcal{S}_j):=\bigcup_{S\in \mathcal{S}_j}\mathcal{E}(S)$,
\begin{equation}
\label{eq: compat}
\begin{gathered}
   \mathcal{S}_1\longrightarrow \mathcal{S}_2 \text{ is in } \mathcal{G}^{\mathrm{sig}}_{\mathfrak{N}}\\ \Downarrow    \\ \exists \text{ directed path from } \mathcal{E}(\mathcal{S}_1) \text{ to } \mathcal{E}(\mathcal{S}_2) \text{ in } \mathcal{G}.
\end{gathered}
\end{equation}

\noindent{\it Fine-graining of quantum networks}| We now introduce the crucial concept of \emph{fine-graining}, which is motivated by physical and practical considerations. For example, when the demand $D$ and price $P$ of a commodity mutually influence each other, this is modelled as a cyclic network in the classical data sciences. However, we know that such a cycle actually represents a physical situation where demand $D_1$ at time $t_1$ influences price $P_2$ at time $t_2>t_1$ which influences demand $D_3$ at time $t_3>t_2$ and so on. Thus, the coarse-grained cyclic network has an explanation in terms of a fine-grained acyclic network, which is why one would not regard the cycle as an exotic physical phenomenon in this case. Here the systems $D$ and $P$ are mapped to a set of systems $\{D_1,D_3,...\}$ and $\{P_2,P_4,...\}$, respectively, under fine-graining.

\begin{figure}
    \centering
    \includegraphics[scale=0.8]{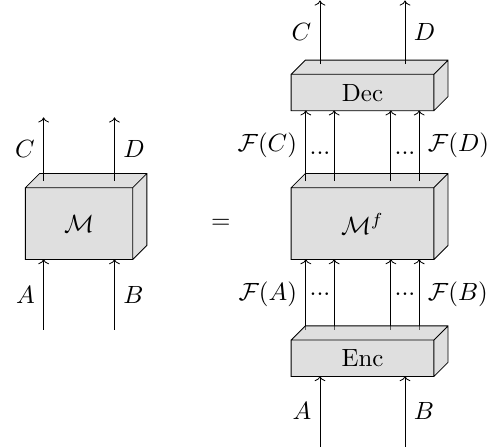}
    \caption{Consider a CPM $\mathcal{M}$ with inputs $A$ and $B$, and outputs $C$ and $D$ and another CPM $\mathcal{M}^f$ such that each in/output of $\mathcal{M}$ maps (through a map $\mathcal{F}$) to a distinct set of in/output systems of $\mathcal{M}^f$. Then we say that $\mathcal{M}^f$ is a fine-graining of $\mathcal{M}$ if there exists a pair $(\mathrm{Enc}, \mathrm{Dec})$ of maps such that (i) the equality in the figure holds and (ii) the signalling relations are preserved under fine-graining, e.g., $\{A\}\rightarrow \{D\} \Rightarrow \mathcal{F}(A)\rightarrow \mathcal{F}(D)$. }
    \label{fig:finegraining}
\end{figure}
%To define this concept, we start with a definition for individual CPMs. 
More formally, Figure~\ref{fig:finegraining} explains when a CPM $\mathcal{M}^f$ can be called a fine-graining of another CPM $\mathcal{M}$ with respect to a fine-graining map~$\mathcal{F}$. Then a network 
$\mathfrak{N}^f$ is defined to be a fine-graining of another network $\mathfrak{N}$ whenever the induced CPM of each sub-network of $\mathfrak{N}^f$ is a fine-graining of the induced CPM of a corresponding sub-network of $\mathfrak{N}$. Further details and examples of composition, signalling, and fine-graining are given in the Supplemental Material. 

\smallskip

\noindent{\it Spacetime realisations and relativistic causality}| To connect the purely information-theoretic concepts discussed thus far to the spatiotemporal picture in a general manner, we model spacetime structure rather minimally, as a partially ordered set $\mathcal{T}$ with order relation $\prec$. This captures the causal structure of spacetime, for instance $P\prec Q$ denotes that the event $P$ is in the past light cone of the event $Q$. We call this a fixed acyclic spacetime.

\begin{figure*}
    \subfloat[]{\includegraphics[scale=1.0]{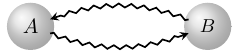}}\qquad\subfloat[]{\includegraphics[scale=1.0]{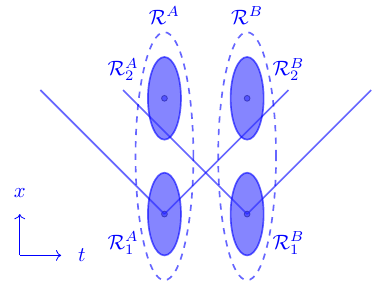}} \qquad  \subfloat[]{\includegraphics[scale=1.0]{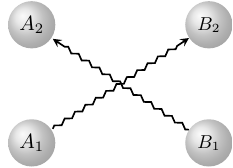}}
 \caption{(a) A cyclic information-theoretic causal structure between systems $A$ and $B$, which can arise in a network that permits bi-directional information flow between them. (b) A cyclic region causal structure between spacetime regions $\cR^A$ and $\cR^B$ (since each region contains an event in the past light-cone of an event in the other). (a) can be compatibly embedded in spacetime by assigning regions $\cR^A$ and $\cR^B$ to $A$ and $B$. The partition $\cR^A=\cR^A_1\cup \cR^A_2$ and $\cR^B=\cR^B_1\cup \cR^B_2$  defines an acyclic region causal structure on 4 (sub-)regions. (c) depicts an acyclic fine-grained causal structure of (a) associated with a spacetime realisation where the nodes $A_1$, $A_2$, $B_1$, $B_2$ are embedded in the sub-regions of (b) (within which more fine-grained interventions can be performed). Our results indicate that the quantum switch experiments have an analogous explanation: the abstract cyclic information-theoretic causal structure unravels to acyclicity when realised in an acyclic spacetime. 
% the spacetime realisation of its cyclic causal structure over two ``nodes'' unravels into an acyclic causal structure over a larger number of ``nodes''. 
}
 \label{fig:main}
\end{figure*}

To realise an information-theoretic network $\mathfrak{N}$ in a fixed spacetime $\mathcal{T}$, we define an embedding $\mathcal{E}$, which assigns to each in/output system $S\in \mathfrak{N}^{\mathrm{sys}}$ a set of spacetime points or \emph{spacetime region} i.e., $\mathcal{E}(S)\in \Sigma(\mathcal{T})$. The image of $\mathcal{E}$ identifies a finite set of spacetime regions, since $\mathfrak{N}^{\mathrm{sys}}$ is finite. We can represent any such set of regions as a directed graph $\mathcal{G}^{\mathrm{reg}}_{\mathcal{T}}$, called the \emph{region causal structure} of a spacetime $\mathcal{T}$. Each node $\mathcal{R}$ of $\mathcal{G}^{\mathrm{reg}}_{\mathcal{T}}$ is a spacetime region $\mathcal{R}\in \Sigma(\mathcal{T})$ and we have a directed edge from $\mathcal{R}_1$ to $\mathcal{R}_2$ 
%$\mathcal{R}_1\xrightarrow[]{R} \mathcal{R}_2$ 
iff $\exists R_1\in\mathcal{R}_1$ and $R_2\in \mathcal{R}_2$ such that $R_1\prec R_2$. %Thus, we can equivalently view a spacetime embedding as assigning to each $S\in \mathfrak{N}^{\mathrm{sys}}$, a node $\mathcal{E}(S)$ of a region causal structure $\mathcal{G}^{\mathrm{reg}}_{\mathcal{T}}$ of a fixed spacetime $\mathcal{T}$. 
Note that the region causal structure can generally be cyclic % with $\mathcal{R}_1\xrightarrow[]{R} \mathcal{R}_2$ and $\mathcal{R}_2\xrightarrow[]{R} \mathcal{R}_1$
(\cref{fig:main}).

Central to our understanding of physical experiments in spacetime is the ability to break down the experiment into processes occurring within smaller spacetime regions. Given a spacetime region $\mathcal{R}$, we can partition it into a set of disjoint regions as $\mathcal{R}=\bigcup_i \mathcal{R}_i$. This may correspond to a fine-graining of a system $S$ embedded in the spacetime region $\mathcal{R}$ as $\mathcal{F}(S)=\{S_i\}_i$, with each fine-grained system $S_i$ embedded in a corresponding fine-grained region $\mathcal{R}_i$. Thus, given a network $\mathfrak{N}$ and a spacetime embedding $\mathcal{E}$ of $\mathfrak{N}$, for every partition of spacetime regions in the image of $\mathcal{E}$, we can consider a corresponding fine-graining of $\mathfrak{N}$. Physically, this captures a minimal requirement that agents (such as experimentalists) can, in principle, perform interventions/physical operations within any of the smaller spacetime regions $\mathcal{R}_i$, up to some resolution of the partition. This resolution can depend on the particular experiment, but for experiments in acyclic spacetimes (i.e., those without closed timelike curves like our physical Minkowski spacetime), the region causal structure over small enough regions $\{\mathcal{R}_i\}_i$ is  acyclic. 

Based on this observation, we define a spacetime realisation of a network $\mathfrak{N}$ in a fixed acyclic spacetime $\mathcal{T}$ as being specified by a spacetime embedding $\mathcal{E}$ of $\mathfrak{N}$ in $\mathcal{T}$, along with at least one fine-graining $\mathfrak{N}^f$ of $\mathfrak{N}$ associated with an acyclic region causal structure $\mathcal{G}^{\mathrm{reg},f}_{\mathcal{T}}$. Crucially, this does not imply that the order of operations in the networks $\mathfrak{N}$ and $\mathfrak{N}^f$ is acyclic, as this has been treated fully independently of the causal order on spacetime regions, so far. To link the two orders, we use the concept of \emph{compatibility} to define a minimal \emph{relativistic causality} condition. We say that a spacetime realisation of a network $\mathfrak{N}$ in a spacetime $\mathcal{T}$ relative to an embedding $\mathcal{E}$ satisfies relativistic causality only if the signalling structure of the fine-graining $\mathfrak{N}^f$ in the specification of the realisation is compatible with a corresponding region causal structure $\mathcal{G}^{\mathrm{reg},f}_{\mathcal{T}}$ (\cref{eq: compat}). This condition ensures the impossibility of agents to signal outside the spacetime's future light cone through fine-grained physical interventions.

\smallskip

\noindent{\it Indefinite causal structures and no-go results}| % for their spacetime realisations
The process matrix framework \cite{Oreshkov2012} is a prominent approach for studying indefinite causal structures \cite{Hardy2005, Chiribella2013, Oreshkov2012}. We consider $N$-party processes where each party (or agent) $A_i$ acts within a local laboratory equipped with a pair of input and output quantum systems, $A_i^I$ and $A_i^O$, and can perform any local quantum operation between these systems. The process matrix $W$ describes the global environment of these labs, telling us how they are connected. Processes where parties act in a fixed acyclic order are called \emph{fixed order processes}, and the ones that do not admit such an explanation (or a convex mixture thereof) are termed \emph{indefinite causal order} processes. We can describe any protocol specified by a process matrix $W$ in terms of a quantum network $\mathfrak{N}_{W,N}$, called the \emph{process network}, by representing $W$ as a completely positive and trace-preserving map (CPTPM) $\cW$, which is composed with the local quantum operations of the $N$ parties through feedback loops defined by the composition operation $\hookrightarrow$ \cite{us_long}. We are now ready to state our two main no-go results. The Supplemental Material provides an intuitive proof sketch of both theorems while highlighting the mathematics underpinning the relevant assumptions.

\begin{theorem}
\label{theorem: nogo1}
For any process $W$, no spacetime realisation of the corresponding process network $\mathfrak{N}_{W,N}$, associated with an embedding $\mathcal{E}$, can simultaneously satisfy the following assumptions:
\begin{enumerate}
    \item $W$ is not a fixed order process.
    \item The spacetime realisation satisfies relativistic causality.
    \item The spacetime region causal structure specified by the image of $\mathcal{E}$ is acyclic.
\end{enumerate}
\end{theorem}

A violation of condition 3 indicates that the agents' in/output systems are not localised in spacetime. The theorem, therefore, shows that physical (not violating relativistic causality) realisations of indefinite causal order processes in a fixed spacetime necessarily involve a non-localisation of systems in spacetime. The next theorem concerns the fine-grained description of such realisations.  %The proof of the second theorem incorporates a natural physical property, that there is a correspondence between in and output spacetime regions associated with each party, which is naturally satisfied in experiments where the device (e.g., wave plate) implementing an agent's operation has a fixed and non-zero input-output processing time. This does not forbid situations where agents receive/send quantum messages at a superposition of different input/output times. 

\begin{theorem}
\label{theorem: nogo2}
Any spacetime realisation of an $N$-party process network $\mathfrak{N}_{W,N}$ that satisfies relativistic causality will admit a fine-grained explanation in terms of a network associated with a fixed order process $W^f$ over $M\geq N$ parties. % RR: I removed the quotation marks, because a reader who only reads the PRL may be unable to understand in which sense they are not parties according to the technical definition.
\end{theorem}
\begin{corollary}
\label{corollary: nogo2}
No spacetime realisation of a process network $\mathfrak{N}_{W,N}$ can simultaneously satisfy the following:
\begin{enumerate}
    \item Relativistic causality.
    \item Absence of any explanation in terms of a process admitting a definite, acyclic (information-theoretic) causal order. 
\end{enumerate}
\end{corollary}

\noindent{\it Implications for experiments}| The \emph{quantum switch} (QS) corresponds to an abstract process $W_{QS}$ which, given any two unitaries $\mathcal{U}^A$ and $\mathcal{V}^B$ (associated with agents Alice and Bob, respectively) implements a quantum superposition of their order of action on a target qubit, depending coherently on the state of a control qubit, 
\begin{equation}
\label{eq: QS}
    (\alpha\ket{0}+\beta \ket{1})^C\ket{\psi}^T \xrightarrow[]{W_{QS}} \alpha \ket{0}^C \mathcal{V}^B\mathcal{U}^A\ket{\psi}^T +  \beta \ket{1}^C \mathcal{U}^A\mathcal{V}^B\ket{\psi}^T
\end{equation}

The QS is incompatible with any explanation in terms of a definite causal ordering between Alice and Bob's operations and is regarded as having an \emph{indefinite (information-theoretic) causal order}. Meanwhile, numerous table-top experiments \cite{Procopio2015, Rubino2017, Goswami2018, Wei_2019, Ho2019, Guo_2020, Goswami_2020, Taddei_2021, Rubino_2021, Felce2021} claim to implement the QS in Minkowski spacetime, a \emph{definite and acyclic spatio-temporal causal structure}. It has been a longstanding debate whether these experiments ``implement'' or ``simulate'' the indefinite causal structure of the QS \cite{Portmann2017,Vilasini_thesis,Paunkovic2019, Oreshkov2019, Ormrod2022, Kabel2024}.%, however a precise definition of what constitutes on or the other has been lacking. 

Our results offer a distinct perspective to this debate at both formal and conceptual levels. %Consider the premise: if an experimental realisation of a process admits a physically meaningful explanation in terms of a definite and acyclic causal order, then it cannot be a fundamental implementation of an indefinite causal order. 
The process network of the QS is associated with a cyclic information-theoretic causal structure between the local labs of Alice and Bob, as it enables bidirectional signalling between them. However, when physically realised in Minkowski spacetime, \cref{theorem: nogo2} guarantees that the realisation will admit a \emph{fine-grained explanation in terms of a fixed and acyclic causal order process $W^f_{QS}$} over a larger number of local labs/operations. This reconciles the apparent tension between the two notions of causality in these scenarios.

If the fine-grained description (which has definite causal order) is operationally verifiable, this provides a concrete criterion to regard such experiments as useful simulations rather than implementations of a fundamentally indefinite causal structure. The fine-grained descriptions involved in our results are indeed operational, for agents can, in principle, perform physical interventions on every system in the description to verify it operationally. The spacetime realization offers access to additional degrees of freedom compared to the process matrix description, allowing for a larger set of interventions. For instance, Alice (or Bob) could perform a different operation $\mathcal{U}^A_1$ or $\mathcal{U}^A_2$ ($\mathcal{V}^B_1$ or $\mathcal{V}^B_2$) depending on the time at which they act, which would lead to a transformation which is not generally possible in the original process matrix of the QS, 
\begin{equation}
\label{eq: QSf}
    (\alpha\ket{0}+\beta \ket{1})^C\ket{\psi}^T \xrightarrow[]{W_{QS}^f} \alpha \ket{0}^C \mathcal{V}^B_2\mathcal{U}^A_1\ket{\psi}^T +  \beta \ket{1}^C \mathcal{U}^A_2\mathcal{V}^B_1\ket{\psi}^T
\end{equation}
The fine-grained process can explain this and reproduces the QS transformation when interventions are restricted as $\mathcal{U}^A_1=\mathcal{U}^A_2$ and $\mathcal{V}^B_1=\mathcal{V}^B_2$, even though it admits a definite and acyclic causal order (see Supplemental Material for details). The fine-grained interventions are crucial for relativistic causality, forbidding signalling outside the future light cone. %This underscores the importance of accounting for all physically allowed interventions, rather than specific performed operations, when analyzing causal structures and their properties such as acyclicity. 
We note that the connection between \cref{eq: QSf} and QS experiments has been previously discussed \cite{Paunkovic2019}, but this example only concerns a specific class of QS realizations. Our no-go theorems and conclusions apply to any process one could attempt to realize in a fixed background spacetime, and any physical realization.

\smallskip

\noindent{\it Discussion and outlook}| We presented two no-go results which reveal fundamental limits on quantum processes realisable in classical background spacetimes, the regime of current experiments. The theorems are derived through a general approach applicable to all indefinite causal order (ICO) processes, all spacetimes where the lightcones define a partial order, and to all experimental realisations of the quantum switch (an ICO process) in Minkowski spacetime \cite{Procopio2015, Rubino2017, Goswami2018, Wei_2019, Ho2019, Guo_2020, Goswami_2020, Taddei_2021, Rubino_2021, Felce2021}, enabling a consistent reconciliation of quantum information-theoretic and relativistic notions of causality.

%The first result underscores that any realization of an ICO process in this regime necessitates the non-localization of systems in spacetime. The second theorem establishes that any quantum experiment in a classical spacetime will inherently admit a fine-grained explanation in terms of a definite and acyclic causal order process, indicating that fundamental implementations of indefinite causal order processes require going beyond this regime. In particular, these conclusions apply to all experimental realisations of the quantum switch (an ICO process) in Minkowski spacetime \cite{Procopio2015, Rubino2017, Goswami2018, Wei_2019, Ho2019, Guo_2020, Goswami_2020, Taddei_2021, Rubino_2021, Felce2021}. Both results follow from minimal assumptions, such as the impossibility of superluminal causation and enable a reconciliation of quantum and relativistic notions of causality. 

%Further, the concept of fine-graining introduced here provides a toolkit to formally investigate how and whether advantages of ICO processes for quantum tasks, found in the abstract information-theoretic approaches translate to physical experiments in spacetime.

The spatiotemporal non-localisation referred to in \cref{theorem: nogo1} can also manifest in realisations of processes with classical mixtures of causal order. Therefore, an important direction for future research is to precisely distinguish between classical vs non-classical forms of information non-localisation in spacetime. Our work provides fine-grained tools and insights for this future direction, while linking it to previous work on causal inequalities. For example, an interesting recent work \cite{Lugt2023} derives correlation bounds satisfied by classical mixtures of fixed order processes under a (coarse-grained) relativistic causality assumption. These are violated by an extended QS protocol, suggesting a device-independent certification of indefinite causal order (at the coarse-grained level). However, our \cref{theorem: nogo2} implies that there cannot be any causal indefiniteness at the fine-grained level in any physical spacetime realisation of such a QS protocol. A fine-grained analysis of this result within our framework shows (consistently with \cref{theorem: nogo1}) that the violation of a causal inequality in such an implementation, if observed, could instead be attributed to non-classicality in the spacetime localisation of the operational events (in/output systems) \cite{Note2}. For a more general and technically detailed discussion on the spatio-temporal interpretation of causal inequalities and the assumptions of the process matrix framework, see Section X of \cite{us_long}. 

A related question is whether and how the advantages of ICO processes for quantum tasks found in the abstract information-theoretic approaches (e.g., \cite{Guerin2016, Chiribella_2021,Chiribella_2012,Zhao_2020,Araujo2014,Guha_2020,mothe2023}) translate to physical experiments in spacetime. This motivates a deeper exploration of whether non-classicality in the spacetime localization of systems (which, as discussed above, can persist despite a definite fine-grained causal order) can serve as a resource for advantages in quantum information processing tasks in spacetime.

%A relevant question to consider for future work is whether advantages of ICO processes for quantum tasks found in the abstract information-theoretic approaches (e.g., \cite{Guerin2016, Chiribella_2021,Chiribella_2012,Zhao_2020,Araujo2014,Guha_2020,mothe2023}) translate to physical experiments in spacetime. Even though such experiments do not display indefinite causality at the fine-grained level (as per our second theorem), can the spatio-temporal non-localization (highlighted by our first theorem) serve as a resource for information processing tasks in spacetime? Noting that such non-localisation is also possible classically, an important future avenue would be to precisely distinguish between classical versus genuinely quantum forms of spacetime non-localization and explore the resource-theoretic aspects of the latter, by analysing the fine-grained causal structure.

Moreover, longstanding challenges persist in characterizing which processes are physically realizable, as several classes of abstract processes are known. Our present work, alongside the companion paper \cite{us_long}, offers a top-down and physically-motivated approach to address this question formally using relativistic principles, which we undertake in follow-up work \cite{Salzger2024, Salzger}. %The concept of fine-graining proves critical here. %Additional no-go results, connections to other causality formalisms and applications of our methods to further research areas such as relativistic quantum information tasks and causal inference are also discussed in the companion paper. 

Finally, our approach of disentangling information-theoretic and spatio-temporal causality notions, coupled with our graph-theoretic proof techniques, could facilitate generalization beyond fixed classical spacetimes and beyond quantum theory (using insights from a related theory-independent approach \cite{VilasiniColbeckPRA, VilasiniColbeckPRL}). 
%Since any quantum experiment in a classical spacetime will inherently admit a fine-grained explanation in terms of a definite and acyclic causal order process (as shown by our second theorem), this indicates that fundamental implementations of indefinite causal order processes require going beyond this regime. 
For future work, this sets the stage for a systematic examination of whether our no-go theorems can be circumvented in quantum-gravitational regimes (where spacetime geometry can be quantum) or post-quantum probabilistic theories. Such endeavours would be relevant for understanding the role of causality and locality in future table-top experiments of quantum gravity \cite{Bose2017, Marletto2017} and can shed light on how such fundamental notions may differ in distinct physical regimes, and the consequences for information processing possibilities therein.

\begin{acknowledgements}
VV's research has been supported by an ETH Postdoctoral Fellowship. VV and
RR acknowledge support from the ETH Zurich Quantum Center, the Swiss National Science Foundation via project No.\ 200021\_188541 and the QuantERA programme via project No.\ 20QT21\_187724.
\end{acknowledgements}

\newpage
%\bibliographystyle{unsrt}
%\bibliography{refs2.bib}

\clearpage

\onecolumngrid
\begin{center}
\large{ {\bf SUPPLEMENTAL MATERIAL}   }
\end{center}

\bigskip

\twocolumngrid

\label{sec: supplement}

\section{Overview of concepts and notation}

\subsection{Information-theoretic networks}
\begin{itemize}
    \item $\mathfrak{N}=(\mathfrak{N}^{maps},\mathfrak{N}^{comp})$: A (possibly cyclic) quantum network defined by a set $\mathfrak{N}^{maps}$ of linear CP maps and a set $\mathfrak{N}^{comp}$ of compositions between pairs of in/output systems of the maps.
    \item $\mathfrak{N}^{sys}$ and $\Sigma(\mathfrak{N}^{sys})$: The set of all in/output systems in a network $\mathfrak{N}$ (up to identification of systems linked by composition), and the powerset (or set of all subsets) of $\mathfrak{N}^{sys}$.
    \item $\mathfrak{N}_{sub}$: a sub-network of a network $\mathfrak{N}$.
    \item $\mathcal{N}_{sub}$: induced map of a sub-network ($\mathfrak{N}_{sub}^{maps}\subseteq \mathfrak{N}^{maps}$, $\mathfrak{N}_{sub}^{comp}\subseteq \mathfrak{N}^{comp}$) $\mathfrak{N}_{sub}$ (obtained by performing the compositions $\mathfrak{N}_{sub}^{comp}$ on the maps in $\mathfrak{N}_{sub}^{maps}$.
    \item $\mathcal{G}^{\mathfrak{N}}_{sig}$: The signalling structure of a network $\mathfrak{N}$.
    \item $\mathfrak{N}^f$: A network which is a fine-graining of a network $\mathfrak{N}$. The maps and compositions of $\mathfrak{N}^f$ are denoted as $\mathfrak{N}^{f,maps}$ and $\mathfrak{N}^{f,comp}$.
    \item $\mathfrak{N}_{W,N}$ and $\mathcal{W}$: Network associated with an $N$-party process matrix $W$, where $\mathcal{W}$ is the CPTP map (process map) of which $W$ is the Choi representation. 

\end{itemize}

\subsection{Spacetime structure}
\begin{itemize}
\item $\mathcal{T}$: spacetime structure (a partially ordered set)
\item $\mathcal{R}$: A spacetime region, $\mathcal{R}\subseteq \mathcal{T}$
\item $\mathcal{G}_{\mathcal{T}}^{reg}$: A region causal structure of a spacetime $\mathcal{T}$.
\item $\mathcal{G}_{\mathcal{T}}^{reg,f}$: A region causal structure which is a fine-graining of $\mathcal{G}_{\mathcal{T}}^{reg}$.
\item $\mathcal{E}$: In the letter, this is mainly used to denote the spacetime embedding which maps each system $S\in \mathfrak{N}^{sys}$ to a spacetime region $\mathcal{E}(S)\subseteq \mathcal{T}$. More generally, it can be used to denote an embedding of systems into any abstract causal structure. 
\end{itemize}
\subsection{Different types of order relations}
\begin{itemize}
    \item $\longrightarrow$: Order relation defined on subsets of systems ($\mathcal{S}\in \Sigma(\mathfrak{N}^{sys})$) that denotes signalling between them. Generally not a partial order.
    \item $\prec$: partial order relation defined on spacetime points $P\in \mathcal{T}$
    \item $\xrightarrow[]{R}$: Order relation defined on spacetime regions $\mathcal{R}\subseteq \mathcal{T}$. Generally not a partial order.
\end{itemize}
\section{Further details of our framework and illustrative examples}

\subsection{Networks and composition}

As explained in the main text, the general information-theoretic structures that we consider, quantum networks, are defined in terms of a set of completely positive maps (CPMs) and a set of compositions specified in terms of a relation $\hookrightarrow$ on the in/output systems of the CPM. Formally, this relation corresponds to \emph{loop composition} \cite{Portmann2017} which is defined as follows.

\begin{definition}[Loop composition \cite{Portmann2017}]
Consider a CPM $\cM$ where $\text{In}(\cM)$ and $\text{Out}(\cM)$ denote the set of all input and output systems of $\cM$ respectively, and let $S\in \text{In}(\cM)$ and $S'\in \text{Out}(\cM)$ have isomorphic Hilbert spaces $\cH^{S}\cong \cH^{S'}$. 
Then we can compose the output subsystem $S$ with the input subsystem $S'$ through loop composition to obtain a CPM $\cM^{S \hookrightarrow S'}$, with $ \text{In}(\cM^{S \hookrightarrow S'})=  \text{In}(\cM)\backslash \{S'\}$ and $ \text{Out}(\cM^{S \hookrightarrow S'})=\text{Out}(\cM)\backslash \{S\}$. The action of $\cM^{S \hookrightarrow S'}$ is given as follows, where we take $\rho^{\text{In}\backslash \{S'\}}$ to be an arbitrary state in $\cL(\text{In}\backslash \{S'\})$,
and treat, for simplicity (and without loss of generality), $S'$ as the last input subsystem.

\begin{equation}
\label{eq: loops}
    \cM^{S \hookrightarrow S'}(\rho^{\text{In}\backslash \{S'\}}):= \sum_{i,j} \bra{i}^{S}\cM\Big(\rho^{\text{In}\backslash \{S'\}}\otimes \ket{i}\bra{j}^{S'}\Big)\ket{j}^{S},
\end{equation}
where the bases appearing in the above equation are taken by convention to be the computational basis of the relevant system. 
\end{definition}

The usual sequential composition of quantum channels can be equivalently expressed in terms of loop composition \cite{Portmann2017, us_long}. For instance, the sequential composition $\cM_2\circ\cM_1$ of a map $\cM_1$ (from $A$ to $B$) followed by a map $\cM_2$ (from $C$ to $D$) with $\cH^{B}\cong \cH^{C}$ is equivalent to loop composing $B\hookrightarrow C$ in $\cM_1\otimes \cM_2$. %This is illustrated in \cref{fig: sequential}.

\subsection{Signalling structure of networks}

Our main results rely on the notion of signalling which is uniquely defined for a given network and can be operationally inferred by agents, given black-box access to the network. Signalling is defined as follows. 
\begin{definition}[Signalling structure of a CPM]
\label{definition: signalling}
We say that there is a \emph{signalling relation} from a subset $\cS_I\subseteq \text{In}(\cM)$ of input systems to a subset $\cS_O\subseteq \text{Out}(\cM)$ of output systems of the map $\cM$ and denote it as $\cS_I\longrightarrow \cS_O$ if there exists a CPTPM $\cN^{\cS_I}:\cL(S_I)\mapsto \cL(\cS_I)$ acting locally on $\cS_I$ such that the following holds.
\begin{equation}
    \label{eq: signalling}
 \tr_{\text{Out}\backslash \cS_O} \circ\cM\neq  \tr_{\text{Out}\backslash \cS_O} \circ\cM\circ \cN^{\cS_I}.
\end{equation}
 The set of all signalling relations of $\cM$ forms a directed graph $\cG_{\cM}^{\text{sig}}$ whose
nodes are elements of the powerset (or sigma algebra) $\Sigma(\text{In}\cup\text{Out})$ of $\text{In}\cup\text{Out}$, with a directed edge $\longrightarrow$ between two nodes whenever there is a signalling relation between those subsets. We refer to this graph as the signalling structure of $\cM$. 
\end{definition}

The reason for defining the signalling structure over the set of sets of systems in the network is that we can have a CPTP map $M$ from an input system $A$ to output systems $B$ and $C$, where $\{A\}$ signals to $\{B,C\}$ but $\{A\}$ does not signal to $\{B\}$ or $\{C\}$ (see \cite{us_long} for an example). However, in certain cases, such as unitary maps \cite{Barrett2020}, signalling relations between pairs of individual systems is sufficient to characterise all possible signalling relations (also between sets of systems) as such situations where a system signals to a set of systems without signalling to subsets thereof are not possible there.

\subsection{Fine-graining}
We provide a more technical overview of the general definition of fine-graining here (Definition 3.17 of \cite{us_long}), as the main text illustrated the special case of a CPM with two inputs and outputs. Let $\mathcal{M}$  and $\mathcal{M}^f$ be two CPMs and let $\mathcal{F}$ be a map that assigns to each input system $I$ from $\mathcal{M}$ a distinct set $\mathcal{F}(I)$ of input systems from $\mathcal{M}^f$ and, similarly, to each output system $O$ from $\mathcal{M}$ a distinct set $\mathcal{F}(O)$ of output systems from $\mathcal{M}^f$. $\mathcal{M}^f$ is called a \emph{fine-graining} of $\mathcal{M}$ with respect to $\mathcal{F}$ if the following two conditions hold: (i)~there exists a pair $(\mathrm{Enc}, \mathrm{Dec})$ of maps\footnote{$\mathrm{Enc}$ is required to be a linear CPTP map and $\mathrm{Dec}$ must be a linear CPTP map on the image of $\mathcal{M}^f \circ \mathrm{Enc}$.} such that $\mathcal{M} = \mathrm{Dec} \circ \mathcal{M}^f \circ \mathrm{Enc}$; (ii)~the signalling relations are preserved, i.e., a set $\cI$ of inputs signals to a set $\mathcal{O}$ of outputs in $\cM$ implies $\cF(\cI)$ signals to $\cF(\mathcal{O})$ in $\cM^f$ (where $\cF(\cI):=\bigcup_{I\in \cI}\cF(I)$ and similarly for $\cF(\mathcal{O})$).

We can then extend this definition to networks: a network $\mathfrak{N}^f$ is a fine-graining of a network $\mathfrak{N}$ relative a systems fine-graining $\mathcal{F}$ mapping each $S\in\mathfrak{N}^{\mathrm{sys}}$ to a distinct set of systems $\mathcal{F}(S)\in \Sigma(\mathfrak{N}^{f,\mathrm{sys}})$, if for every sub-network of $\mathfrak{N}$, there is a corresponding sub-network of $\mathfrak{N}^f$ such that the induced map of the latter is a fine-graining of the induced map of the former.

Figure 2 of the main text illustrates the concept of fine-graining at the level of graphs. In this Supplement, we illustrate two networks whose information-theoretic causal structure would correspond to graphs of Figure 2 (a) and Figure 2 (c) respectively, such that the latter network is a fine-graining of the former. %For more detailed examples, which make explicit the descriptions of the coarse and fine-grained networks along with the encoding-decoding schemes, we refer the reader to Section 3.4 and Appendix C of \cite{us_long}. 

\section{Indefinite causal order processes}
\subsection{The process matrix framework}

Multiple frameworks \cite{Hardy2005, Chiribella2013, Oreshkov2012} have formalised the concept of indefinite causal order (ICO) processes. Here 
we focus on the process matrix framework \cite{Oreshkov2012}, and refer the reader to \cite{us_long} for details on how our results apply in the equivalent framework of higher-order quantum maps \cite{Chiribella2013}.

%both of which can equivalently describe abstract quantum information protocols where the order of operations may no longer be definite and acyclic. %, here we focus on the process matrix framework as questions relating to the physicality of ICOs and different classes of ICOs have often been discussed here. We nevertheless outline the connection to the higher-order point of view, as this is also useful.

The process matrix framework \cite{Oreshkov2012} describes multi-party protocols involving a finite set $\{A_1,..,A_N\}$ of agents, where each agent is associated with a local quantum laboratory having a quantum input system $A_i^I$ and quantum output system $A_i^I$. The most general operation that $A_i$ can perform between these quantum in and output systems is given by a quantum instrument $\mathcal{J}_{a_i}$, parametrised by a choice of classical setting $a_i$. $A_i$'s instrument for each setting $a_i$ is a set of CPMs from $A_i^I$ to $A_i^O$, $\mathcal{J}_{a_i}:=\{\cM^{A_i}_{x_i|a_i}\}_{x_i\in X_i}$ one for each outcome $x_i\in X_i$. Moreover, $\sum_{x_i}\cM^{A_i}_{x_i|a_i}$ is CPTP. The process matrix $W$ is an operator living in the joint state space of all in and outputs $A^1_I,A^1_O,...,A^N_I,A^N_O$ which describes the global environment of these labs. Given a set of local CPMs, one for each party $\cM^{A_1}_{x_1|a_1},...,\cM^{A_N}_{x_N|a_N}$, the process matrix $W$ enables us to compute the joint probability of the outcomes given the settings through the following rule
\begin{equation}
    P(x_1,...,x_N|a_1,...,a_N)=\tr(\bigotimes_iM^{A_i^IA_i^O}_{x_i|a_i}. W),
\end{equation}

where $M^{A_i^IA_i^O}_{x_i|a_i}$ (living in the joint space of $A_i^I$ and $A_i^O$) is the Choi-Jamiolkowski representation of the CPM $\cM^{A_i}_{x_i|a_i}$. Importantly, a valid process matrix yields valid, normalised probabilities through the above rule, for every choice of local operations.

Processes compatible with a fixed acyclic ordering between the agents are called \emph{fixed order processes}. We illustrate this for the bipartite case. Consider a bipartite process matrix $W$ involving agents Alice and Bob. This is said to be compatible with the fixed order Alice before Bob, and denoted as $W^{A\prec B}$ if Alice's outcome is independent of Bob's setting, i.e. for all choices of local operations we get probabilities satisfying $P(x|ab)=P(x|a)$. Likewise for the fixed order process $W^{B\prec A}$, we have $P(y|ab)=P(y|b)$. If a process $W$ can be expressed as $W=qW^{A\prec B}+(1-q)W^{B\prec A}$ for some $q\in[0,1]$, then $W$ is called causally separable and otherwise it is said to be causally non-separable. Causally non-separable processes are regarded as indefinite causal structures as they do not admit an explanation in terms of any fixed order process or convex mixtures thereof. For the general multi-partite definition of these and related concepts, we refer the reader to \cite{Oreshkov2016, Branciard2016}.

%One can equivalently describe the action of the process on the local operations as a higher-order transformation which maps the local operations of the $N$ agents to the corresponding probabilities \cite{Chiribella2013}. Generally higher-order transformations describe transformations on quantum channels (of which the local operations described above and probability distribution are examples), as opposed to standard quantum transformation which act on states. Our results can equivalently be applied in the process matrix and the higher-order respresentations, as detailed in the associated paper \cite{us_long}. %As detailed in the associated paper \cite{us_long} (see also \cite{Araujo2016, Araujo2017}), we can represent the process matrix/higher-order transformation as a single CPTP map $\mathcal{W}$ whose inputs are the outputs $A_O^1,...,A_O^N$ of the agents, and whose outputs are the inputs $A_I^1,...,A_I^N$ of the agents, where $\mathcal{W}$ acts on the local operations through loop composition, thus allowing these objects to be defined in terms of the cyclic quantum networks introduced in our work. 

\begin{figure*}
    \centering
    \subfloat[]{
    \begin{tikzpicture}[line width=0.20mm, scale=0.8, transform shape,trace/.pic={\draw [thick](-0.4,0)--(0.4,0);\draw [thick](-0.3,0.1)--(0.3,0.1);\draw [thick](-0.2,0.2)--(0.2,0.2);\draw [thick](-0.1,0.3)--(0.1,0.3);}]
	\node[cube, ,draw=black,fill=lightgray, fill opacity=0.5, text opacity=1,
  minimum width=1.5cm,minimum height=1.5cm] at (0,0){$\cM_1$};
\node[cube, ,draw=black,fill=lightgray, fill opacity=0.5, text opacity=1,
  minimum width=1.5cm,minimum height=1.5cm] at (0,3){$\cM_2$};

\draw[thick, ->] (0,0.85)--node[anchor=east]{$B$}(0,2.25);
\draw[thick, ->] (0,3.85)--node[anchor=east]{$A$}(0,4.85);
\draw[thick, ->] (0,-1.75)--node[anchor=east]{$A$}(0,-0.75);
\draw[thick] (2,-1.75)--(2,4.85);
\draw[thick] (0,-1.75) to[out=270,in=270] (2,-1.75);
\draw[thick] (0,4.85) to[out=90,in=90] (2,4.85);

\end{tikzpicture}}\qquad\qquad\qquad
\subfloat[]{\begin{tikzpicture}[line width=0.20mm, scale=0.8, transform shape,trace/.pic={\draw [thick](-0.4,0)--(0.4,0);\draw [thick](-0.3,0.1)--(0.3,0.1);\draw [thick](-0.2,0.2)--(0.2,0.2);\draw [thick](-0.1,0.3)--(0.1,0.3);}]
\begin{scope}[shift={(0,2)}]
    \node[cube, ,draw=black,fill=lightgray, fill opacity=0.5, text opacity=1,
  minimum width=4cm,minimum height=2cm] at (0,0){};

\begin{scope}[shift={(-1,0.8)}]
     \clip (-0.6,0) rectangle (0.6,-0.6);
    \draw[fill=gray, fill opacity=0.3] (0,0) circle(0.6);
    \draw (-0.6,0) -- (0.6,0);
    \end{scope}
\node at (1.5,0.5) {\large{$\cM_2^f$}};
    \draw (1,-0.8) pic {trace};
  \draw[thick, ->] (-1,-2.5)--node[anchor=east]{$B_1$}(-1,-1);
  \draw[thick] (-1,-2.7)--(-1,-2.5);
  \draw[thick] (1,-2.5)--node[anchor=east]{$B_2$}(1,-1);
  \draw[thick,->] (1,-1)--(1,-0.8);
  \draw[thick, ->] (-1,1)--node[anchor=east]{$A_1$}(-1,2.5);
  \draw[thick] (-1,0.8)--(-1,1);
  \draw[thick, ->] (1,1)--node[anchor=east]{$A_2$}(1,2.5);
  \draw[thick] (-1,-1) to[out=90,in=270] (1,1);
\end{scope}

\begin{scope}[shift={(0,-1.5)}]
    \node[cube, ,draw=black,fill=lightgray, fill opacity=0.5, text opacity=1,
  minimum width=4cm,minimum height=2cm] at (0,0){};

\begin{scope}[shift={(-1,0.8)}]
     \clip (-0.6,0) rectangle (0.6,-0.6);
    \draw[fill=gray, fill opacity=0.3] (0,0) circle(0.6);
    \draw (-0.6,0) -- (0.6,0);
    \end{scope}
\node at (1.5,0.5) {\large{$\cM_1^f$}};
    \draw (1,-0.8) pic {trace};
  \draw[thick, ->] (-1,-2.5)--node[anchor=east]{$A_1$}(-1,-1);
  \draw[thick] (1,-2.5)--node[anchor=east]{$A_2$}(1,-1);
  \draw[thick,->] (1,-1)--(1,-0.8);
  \draw[thick] (-1,-1) to[out=90,in=270] (1,1);
\end{scope}
\draw[thick] (-3,-4)--(-3,4.5);
\draw[thick] (3,-4)--(3,4.5);
\draw[thick] (-3,4.5) to[out=90,in=90] (-1,4.5);
\draw[thick] (1,4.5) to[out=90,in=90] (3,4.5);
\draw[thick] (-3,-4) to[out=270,in=270] (-1,-4);
\draw[thick] (1,-4) to[out=270,in=270] (3,-4);
    \end{tikzpicture}}
    \caption{(a) A cyclic quantum network, where for suitably chosen CPMs $\cM_1$ and $\cM_2$, the information-theoretic causal structure would be that of Figure 2(a) from the main text, where $A\protect\longrsquigarrow B$ and $B\protect\longrsquigarrow A$. (b) Although~(a) is cyclic, it is possible that there exists a network, such as the one depicted here, that is a fine-graining of~(a), where the systems $A$ and $B$ map to the sets $\cF(A)=\{A_1,A_2\}$ and $\cF(B)=\{B_1,B_2\}$ of systems and where $\cM_1^f$ and $\cM_2^f$ are fine-grainings of the CPMs $\cM_1$ and $\cM_2$, respectively. The semicircles denote state preparations and \ground\ denotes trace. The fine-grained network clearly has an acyclic causal structure, that of Figure 2(c) of the main text, where $A_1\protect\longrsquigarrow B_2$ and $B_1\protect\longrsquigarrow A_2$ are the only relations. See Appendix~C and~G of \cite{us_long} for explicit examples of cyclic networks that admit such acyclic fine-grainings. The quantum switch process is one such example. %The process has a cyclic causal structure but its spacetime realisation has an acyclic fine-grained causal structure as suggested by our Theorem 2.
    }
    \label{fig: fg_network}
\end{figure*}
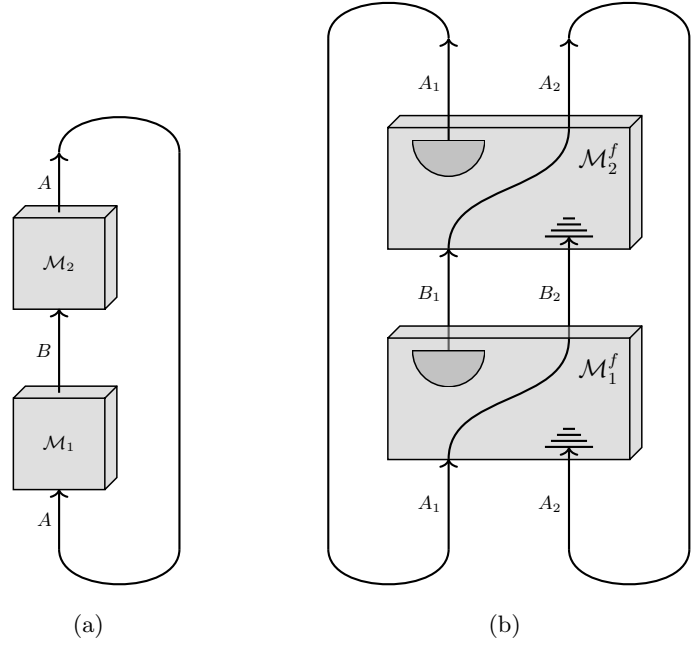
\subsection{The quantum switch process}

The quantum switch is one of the simplest examples of an indefinite causal order process and has attracted much interest in the community. It can be described as a 4-partite process matrix $W_{QS}$ involving agents $A$, $B$, $C$ and $D$, where $C$ acts in the global past of all others and $D$ in the global future. Thus the input $C^I$ and output $D^O$ can be taken to be trivial, 1 dimensional systems. The output of $C$ and input of $D$ include the control and target degrees of freedom, $C^O:=C^O_C\otimes C^O_T$ and $D^I=D^I_C\otimes D^I_T$ with the control being a qubit and target a qudit. The process matrix is given by $W_{QS}=\ket{W_{QS}}\bra{W_{QS}}$ where $\ket{W_{QS}}$ is
\begin{align}
\label{eq: w'qs}
\begin{split}
\ket{W^{QS}}=& |\mathds{1}\rrangle^{C^O_TA^I}|\mathds{1}\rrangle^{A^OB^I}|\mathds{1}\rrangle^{B^OD^I_T}\ket{00}^{C^O_CD^I_C}\\
&+|\mathds{1}\rrangle^{C^O_TB^I}|\mathds{1}\rrangle^{B^OA^I}|\mathds{1}\rrangle^{A^OD^I_T}\ket{11}^{C^O_CD^I_C}  
\end{split}
\end{align}

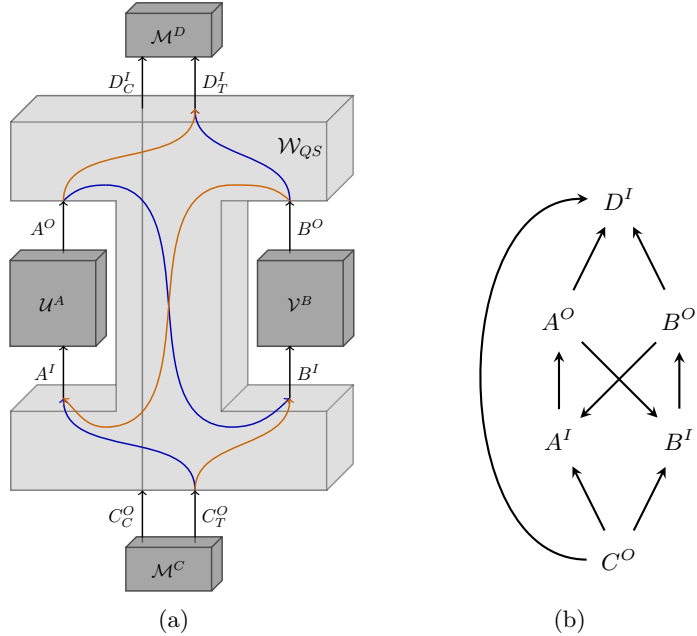
\begin{figure*}
\centering
\subfloat[]{
\centering
   \includegraphics[scale=0.7]{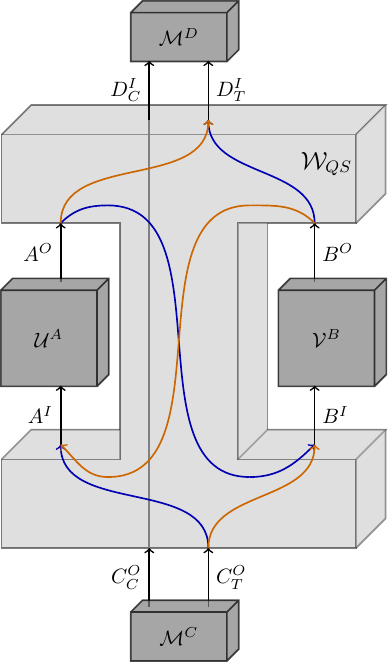}   
}\qquad\quad
\subfloat[]{
\centering
 \begin{tikzpicture}[scale=0.8]
%\node[shape=circle] (OC) at (0,0) {$C^O_C$};
\node[shape=circle] (OT) at (3,0) {$C^O$}; \node[shape=circle] (AI2) at (2,2) {$A^I$}; \node[shape=circle] (BI2) at (4,2) {$B^I$}; \node[shape=circle] (AO3) at (2,4) {$A^O$}; \node[shape=circle] (BO3) at (4,4) {$B^O$};
 %\node[shape=circle] (IC) at (0,6) {$D^I_C$};
 \node[shape=circle] (IT) at (3,6) {$D^I$};
% \draw[arrows={-stealth}, thick] (OC)--(IT);\draw[arrows={-stealth}, thick] (OC)--(AI2); \draw[arrows={-stealth}, thick] (OC)--(BI2); \draw[arrows={-stealth}, thick] (OC)--(IC);
\draw[arrows={-stealth}, thick] (OT)--(AI2); \draw[arrows={-stealth}, thick] (OT)--(BI2); \draw[arrows={-stealth}, thick] (AO3)--(BI2); \draw[arrows={-stealth}, thick] (BO3)--(AI2); \draw[arrows={-stealth}, thick] (AO3)--(IT); \draw[arrows={-stealth}, thick] (BO3)--(IT); \draw[arrows={-stealth}, thick] (AI2)--(AO3);  \draw[arrows={-stealth},  thick] (BI2)--(BO3);  
\path[arrows={-stealth}, thick] (OT)edge[bend left=90](IT);
\end{tikzpicture}  }
 
	\caption{(a) The quantum switch process. Agent $C$ prepares an initial state of the control and target (left side of \cref{eq: QS}). The target takes the blue path (going first to $A$ and then $B$) or the orange path (going first to $B$ and then $A$) depending coherently on the state of the control being $\ket{0}$ or $\ket{1}$. (b) Information-theoretic causal structure of the quantum switch. We have represented the causal structure with the arrows $\protect\longrightarrow$ rather than $\protect\longrsquigarrow$ here as causal relations and signalling relations (on individual systems) coincide in this example, and there are no new non-trivial signalling relations over sets of systems. This is generally not the case for CPTPMs, but true for unitaries \cite{Barrett2020}. The quantum switch process in (a) corresponds to a network of unitary channels as the process matrix $W_{QS}$ is the Choi representation of a unitary channel $W_{QS}$ from $C^O_C$, $C^O_T$, $A^O$, $B^O$ to $A^I$, $B^I$, $D^I_C$, $D^I_T$, and the local operations are unitaries as well. The causal structure is cyclic since Alice and Bob can signal to each other in both directions.  }
	\label{fig: QS}
\end{figure*}

This process matrix implements a coherence controlled superposition of the orders $C\prec A\prec B\prec D$ (when control is $\ket{0}$) and $C\prec B\prec A\prec D$ (when control is $\ket{1}$) and is depicted in \cref{fig: QS}. It can be shown that there exists no $q\in [0,1]$ and fixed order processes $W^{C\prec A\prec B\prec D}$ and $W^{C\prec B\prec A\prec D}$ such that $W_{QS}= qW^{C\prec A\prec B\prec D}+(1-q)W^{C\prec B\prec A\prec D}$ i.e., $W_{QS}$ is causally non-separable. Given any unitary local operations $\mathcal{U}^A: A^I\mapsto A^O$ and $\mathcal{V}^B: B^I\mapsto B^O$ for $A$ and $B$, $W_{QS}$ implements the following transformation (same as Equation from main text upto labelling of systems) on the control and target going from global past to future
\begin{align}
\label{eq: QS}
\begin{split}
     (\alpha\ket{0}+\beta \ket{1})^{C^O_C}\ket{\psi}^{C^O_T} \xrightarrow[]{W_{QS}} &\alpha \ket{0}^{D^I_C} \mathcal{V}^B\mathcal{U}^A\ket{\psi}^{D^I_T}\\ &+  \beta \ket{1}^{D^I_C} \mathcal{U}^A\mathcal{V}^B\ket{\psi}^{D^I_T}   
\end{split}
\end{align}

\section{Further details on main results}

In this section, we provide an intuitive proof sketch of the two main theorems stated in the main text, which correspond to Theorems 8.1 and 8.6 from the longer companion paper \cite{us_long}. The full proof is provided in \cite{us_long} as it requires further technical details and intermediate results to be rigorously formalised. 

The action of a process $W\in A^I_1\otimes A^I_1\otimes...\otimes A_N^I\otimes A_N^O$ on the local operations of the $N$ agents can be viewed as a (possibly cyclic) quantum network $\mathfrak{N}_W$ formed by the composition of a CPTPM $\cW$ (the \emph{process map}) from inputs $A_1^O$,...,$A^O_N$ (outputs of the parties) to outputs $A_I^1$,...,$A^I_N$ (inputs of the parties) with the local operations of the parties \cite{Araujo2016, us_long}. $\mathfrak{N}_W$, which is called the process network, is the central object in these theorems, as we have seen in the main text.

\subsection{Proof sketch of Theorem 1}

\setcounter{theorem}{0}
\begin{theorem}
%\label{theorem: nogo1}
For any process $W$, no spacetime realisation of the corresponding process network $\mathfrak{N}_W$ (associated with an embedding $\mathcal{E}$) can simultaneously satisfy the following assumptions:
\begin{enumerate}
    \item $W$ is not a fixed order process.
    \item The spacetime realisation satisfies relativistic causality.
    \item The spacetime region causal structure specified by the image of $\mathcal{E}$ is acyclic.
\end{enumerate}
\end{theorem} 
{\it Proof sketch}| The main step in the proof is an intermediate result that we prove in \cite{us_long} (Theorem 7.4), which states that a process matrix $W$ is not a fixed order process if and only if the compatibility of the signalling structure of the corresponding network $\mathfrak{N}_{W,N}$ with a directed graph $\cG$ certifies the existence of a directed cycle in $\cG$. The next step is to recall that once we embed $\mathfrak{N}_{W,N}$ in a spacetime through an embedding $\cE$ that maps each system in the network to a node of a region causal structure $\cG^{\mathrm{reg}}_{\cT}$, relativistic causality requires the signalling structure of $\mathfrak{N}_{W,N}$ to be compatible with the directed graph $\cG^{\mathrm{reg}}_{\cT}$. It follows that imposing condition 1. and condition 2. of the theorem implies the existence of a directed cycle in $\cG^{\mathrm{reg}}_{\cT}$, which is equivalent to a violation of condition 3., and establishes the theorem.

\subsection{Proof sketch of Theorem 2}
The proof of this theorem incorporates a natural physical requirement, which we call input-output correspondence. This is satisfied in experiments where the device (e.g., wave plate) implementing an agent's operation has a fixed and non-zero input-output processing time, and does not forbid situations where agents receive/send quantum messages at a superposition of different input/output times. 

We describe this property more formally before sketching the proof of the theorem. This condition ensures that the spacetime regions $\cR^{A^I}$ and $\cR^{A^O}$ assigned to the input and output systems $A^I$ and $A^O$ of each agent $A$ are fine-grained in a similar manner. Formally, for every fine-graining of $\cR^{A^I}$ into $n$ subregions $\cP_1,..,\cP_n$ i.e, $\cR^{A^I}=\bigcup_{i=1}^n\cP_i$, we consider a corresponding partition of  $\cR^{A^O}$ into $n$ subregions $\cQ_1,..,\cQ_n$ i.e, $\cR^{A^O}=\bigcup_{i=1}^n\cQ_i$ such that $\cP_i\xrightarrow[]{R} \cQ_i$ ($\cP_i$ precedes $\cQ_i$ in the region causal structure). Then we say that the regions $\cR^{A^I}$ and $\cR^{A^O}$ have a cardinality $n$ relative to this partition. This condition would be physically satisfied for instance if each agent's device has a fixed input-output processing time $\Delta t$. Then if $A$'s lab is located at $\vec{r}_A$ and can receive inputs at times $t_1,...,t_n$, we have $\cR^{A^I}=\{(\vec{r}_A,t_1),..., (\vec{r}_A,t_n)\}$, and $\cR^{A^I}=\{(\vec{r}_A,t_1+\Delta t),..., (\vec{r}_A,t_n+\Delta t)\}$. Here the sub-regions $\cP_i$ and $\cQ_i$ are individual spacetime points and $\xrightarrow[]{R}$ reduces to the spacetime order relation $\prec$, clearly we have $(\vec{r}_A,t_i)\prec (\vec{r}_A,t_i+\Delta t)$ for all $i$ and the input-output correspondence is satisfied.

\begin{figure*}
\centering
\subfloat[]{% 
\includegraphics[scale=0.6]{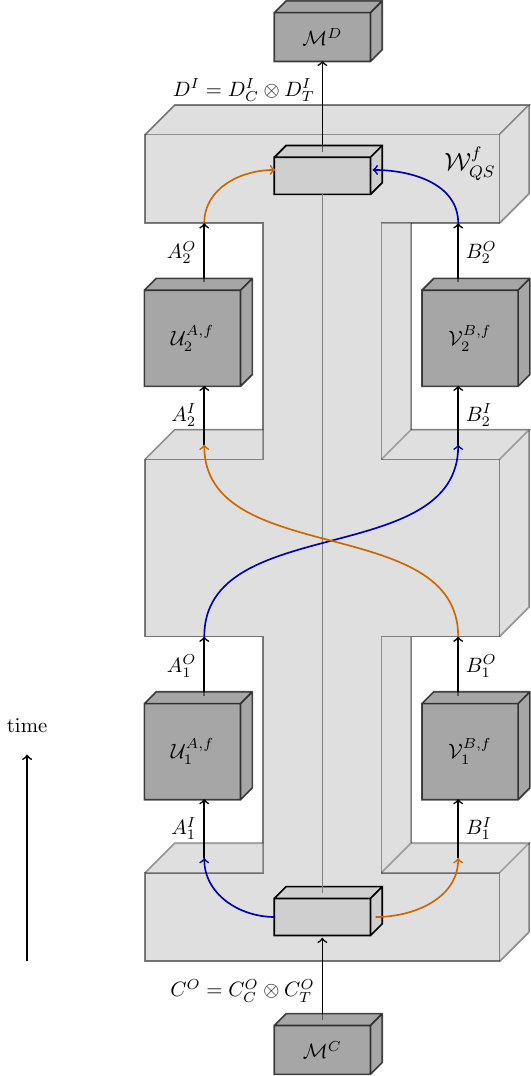}}%
\qquad\qquad\qquad\subfloat[]{%
\begin{tikzpicture}[scale=0.8]
\node[shape=circle] (OT) at (3,0) {$C^O$}; \node[shape=circle] (AI2) at (1,2) {$A^I_1$}; \node[shape=circle] (BI2) at (5,2) {$B^I_1$}; \node[shape=circle] (AO3) at (1,4) {$A^O_1$}; \node[shape=circle] (BO3) at (5,4) {$B^O_1$};
\node[shape=circle] (AI4) at (1,6) {$A^I_2$}; \node[shape=circle] (BI4) at (5,6) {$B^I_2$};\node[shape=circle] (AO5) at (1,8) {$A^O_2$}; \node[shape=circle] (BO5) at (5,8) {$B^O_2$}; 
\node[shape=circle] (IT) at (3,10) {$D^I$}; 
\draw[arrows={-stealth}, thick] (OT)--(AI2); \draw[arrows={-stealth}, thick] (OT)--(BI2); \draw[arrows={-stealth}, thick] (AO3)--(BI4); \draw[arrows={-stealth}, thick] (BO3)--(AI4); \draw[arrows={-stealth}, thick] (AO5)--(IT); \draw[arrows={-stealth}, thick] (BO5)--(IT); \path[arrows={-stealth}, thick] (OT)edge[bend left=90](IT);
\draw[arrows={-stealth}, thick] (AI2)--(AO3); \draw[arrows={-stealth}, thick] (AI4)--(AO5); \draw[arrows={-stealth}, thick] (BI2)--(BO3); \draw[arrows={-stealth}, thick] (BI4)--(BO5);   
\end{tikzpicture}}%
\caption{ Fine-graining of QS associated with its spacetime realisation: (a) A schematic of the fine-grained network associated with typical spacetime realisations of QS. It is associated with a 6-partite process $\mathcal{W}_{QS}^f$ which is a fixed order process but implements, for arbitrary unitaries $\mathcal{U}^A$ and $\mathcal{V}^B$, the QS transformation of \cref{eq: QS} from the control and target received on $C_O$ to corresponding systems in $D^I$ when Alice and Bob (who now act at two distinct times $t_1$ and $t_2>t_1$) apply the same unitary at both times. (b) The causal structure of the fine-grained network, which is definite and acyclic. Here as well (similar to \cref{fig: QS}) we represent the causal structure with solid arrows as the causal relations and signalling relations coincide.}
	\label{fig: PMQS_fg}
\end{figure*}
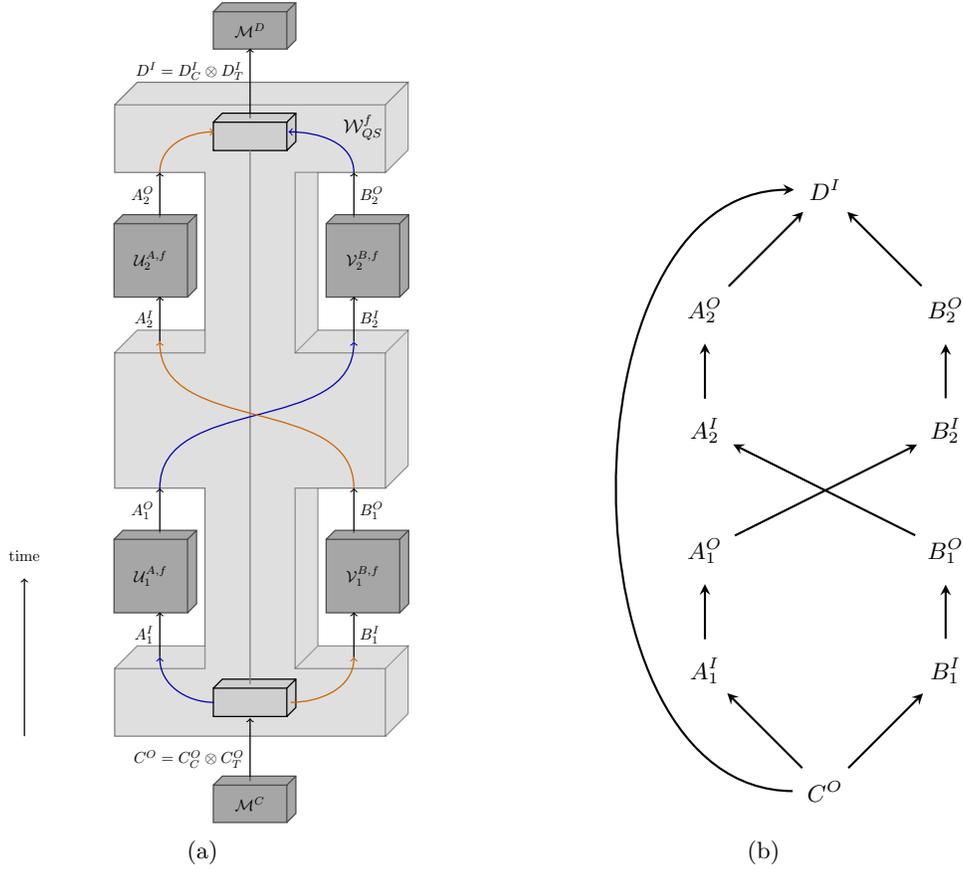
\setcounter{theorem}{1}
\begin{theorem}
%\label{theorem: nogo2}
Any spacetime realisation of a process network $\mathfrak{N}_{W,N}$ associated with any $N$-party process $W$ that involves an input-output correspondence and satisfies relativistic causality, will admit a fine-grained explanation in terms of a fixed order process $W^f$ over a larger number $M\geq N$ of possibly communicating local labs/``parties''.
\end{theorem}

We note that in the above statement of the theorem, as compared to the version in the main text, we have made more explicit the fact that the local labs in the fine-grained description can be potentially communicating, i.e., can generally have communication channels through ancilliary systems, outside of those encoded within the process.

{\it Proof sketch}| The proof can be broken down into the following three steps.

{\bf 1. Fine-graining and relativistic causality} Recall that a spacetime realisation of a network $\mathfrak{N}_{W,N}$ is specified by a spacetime embedding $\cE$ of it into a region causal structure $\cG^{\mathrm{reg}}_{\cT}$ along with a fine-graining $\mathfrak{N}_{W,N}^f$ of $\mathfrak{N}_{W,N}$ defined relative to a partitioning of nodes of $\cG^{\mathrm{reg}}_{\cT}$. Specifically, the partitioning is required to lead to an acyclic region causal structure $\cG^{\mathrm{reg},f}_{\cT}$ over the sub-regions in the partition, capturing the essence of the fact that we are realising the network in an acyclic background spacetime. Relativistic causality imposes the compatibility between the fine-grained signalling structure i.e., the signalling structure of $\mathfrak{N}_{W,N}^f$ with the acyclic region causal structure $\cG^{\mathrm{reg},f}_{\cT}$. This ensures that even fine-grained interventions (which are physically possible) in the spacetime realisation cannot lead to signalling outside the spacetime's future light cone. 

{\bf 2. Fine-grained network as a process network} In the next step of the proof, we invoke the input-output correspondence which allows us to suppose that for each party $A_i$, the in/output regions $\cR^{A^I_i}$ and $\cR^{A^O_i}$ have the same cardinality $n_i$ relative to the partition defining the region causal structure $\cG^{\mathrm{reg},f}_{\cT}$. The fine-grained network will then, for each party $A_i$, have $n_i$ fine-grained input systems corresponding to the coarse-grained input $A^I_i$ and $n_i$ fine-grained output systems corresponding to the coarse-grained output $A^O_i$. It can be shown that $\mathfrak{N}_{W,N}^f$ corresponds to the composition of a CPTP map $\cW^f$ from all the fine-grained outputs ($\sum_{i=1}^Nn_i$ of them) to the fine-grained inputs ($\sum_{i=1}^Nn_i$ of them), together with local maps of $M=\sum_{i=1}^Nn_i$ ``fine-grained'' parties. These parties could be communicating, in the sense that the local operations of two parties need not be in tensor product form, but can have a memory connecting them. For example, in \cref{fig: PMQS_fg}, we have a tensor product form between the local operations (in dark gray) of the two Alices and two Bobs, by allowing an additional output $O$ to say $\mathcal{U}^{A,f}_1$ and additional input $I$ to $\mathcal{U}^{A,f}_2$ and including $O\hookleftarrow I$, we allow for communication between the two Alices. Altogether, this argument establishes that the fine-grained network $\mathfrak{N}_{W,N}^f$ is a process network associated with the $M\geq N$ party process map $\cW^f$.

{\bf 3. Fine-grained process has a fixed order} The final step is to show that the fine-grained process $\cW^f$ is a fixed order process i.e., has a well-defined acyclic causal order according to the process matrix framework. This immediately follows from Theorem 1 applied to the fine-grained process network $\mathfrak{N}_{W,N}^f$, since it satisfies relativistic causality and is embedded in an acyclic region causal structure $\cG^{\mathrm{reg},f}_{\cT}$. Hence, conditions 2. and 3. of Theorem 1 are satisfied which implies that condition 1. is violated, i.e., $\cW^f$ must be a fixed order process.

\subsection{Fine-graining of the quantum switch}
\cref{fig: PMQS_fg} illustrates the fine-grained process network corresponding to a spacetime realisation of the quantum switch process. As expected from Theorem 2, this network corresponds to a fixed and acyclic causal order process $\cW^f_{QS}$, where each in/output system $S\in \{A^I,A^O,B^I,B^O\}$ is fine-grained to two systems $S_1$ and $S_2$ where $S_1$ is assigned an earlier spacetime region than $S_2$. This fine-graining in fact corresponds to the description of the quantum switch previously presented in the causal box framework \cite{Portmann2017}. Here, when the control is $\ket{0}$, the target qudit arrives to Alice at time $t_1$ and Bob at time $t_2>t_1$ (blue path) and a vacuum state $\ket{\Omega}$ (representing the absence of a physical system) occupies the other path (orange path), and when the control is $\ket{1}$ the target qudit arrives to Bob at time $t_1$ and Alice at time $t_2>t_1$ (orange path) and a vacuum state $\ket{\Omega}$ occupies the other path (blue path). Thus, while the coarse-grained systems $S\in \{A^I,A^O,B^I,B^O\}$ were $d$ dimensional (qudit spaces), the fine-grained systems $S_1$ and $S_2$ are each $d+1$ dimensional as they include the 1-dimensional vacuum state $\ket{\Omega}$ which is orthogonal to the remaining states. The fine-grained local operations act identically to the coarse-grained operations on the non-vacuum (qudit) subspace and leave the vacuum subspace invariant (see \cite{us_long} for details). Importantly, unlike the coarse-grained QS process, the fine-grained network can explain the transformation in Equation (3) of the main text which involves 4 distinct unitary operations, but nevertheless recovers the QS transformation of \cref{eq: QS} when restricting these operations to satisfy $\mathcal{U}_1^{A,f}=\mathcal{U}_2^{A,f}$ and $\mathcal{U}_1^{B,f}=\mathcal{U}_2^{B,f}$. %This highlights that for understanding the causal structure (according to both notions), studying the overall transformation, such as \cref{eq: QS} alone is insufficient. Both the fine-grained process (which admits a fixed and acyclic causal order) and the original QS process (which admits an indefinite causal order) can reproduce this transformation. Rather, our work stresses the importance of taking into account the set of all physically possible interventions in a given regime, such as situations where $\mathcal{U}_1^{A,f}\neq \mathcal{U}_2^{A,f}$ (which is possible in Minkowski spacetime, by choosing a time-dependant unitary), in making conclusions regarding properties of the causal structures, such as the acyclicity.

We note that ,any previous works have suggested a similar description of the quantum switch experiments, in terms of vacuum states. Our theorems apply more generally and do not rely on the existence of such a vacuum state and neither do they require perfect correlations between the control and the time of arrival of the qudit which exists in this example. 

\end{document}